\documentclass[conference]{IEEEtran}%
\usepackage{amsfonts}
\usepackage{amsmath}
\usepackage{amssymb}
\usepackage{graphicx}%
\newtheorem{theorem}{Theorem}

\newtheorem{proposition}[theorem]{Proposition}
\newtheorem{remark}[theorem]{Remark}

\begin{document}

\title{A General Coding Scheme for Two-User Fading Interference Channels }

%

%TCIMACRO{\TeXButton{Author Information}{\author{\authorblockN
%{Lalitha Sankar\authorrefmark{1}, Elza Erkip\authorrefmark{2}%
%, H. Vincent Poor\authorrefmark{1}}
%\authorblockA{\authorrefmark{1}Dept. of Electrical Engineering,
%Princeton University,
%Princeton, NJ 08544.
%{lalitha,poor}@princeton.edu\\}
%\authorblockA{\authorrefmark{2}Dept. of Electrical and Computer Engineering,
%Polytechnic University,
%Brooklyn, NY 11201.
%elza@poly.edu\\}}}}%
%BeginExpansion
\author{\authorblockN{Lalitha Sankar\authorrefmark{1}%
, Elza Erkip\authorrefmark{2}, H. Vincent Poor\authorrefmark{1}}
\authorblockA{\authorrefmark{1}Dept. of Electrical Engineering,
Princeton University,
Princeton, NJ 08544.
{lalitha,poor}@princeton.edu\\}
\authorblockA{\authorrefmark{2}Dept. of Electrical and Computer Engineering,
Polytechnic University,
Brooklyn, NY 11201.
elza@poly.edu\\}}%
%EndExpansion
%

%TCIMACRO{\TeXButton{Make Title}{\maketitle}}%
%BeginExpansion
\maketitle
%EndExpansion
%

%TCIMACRO{\TeXButton{Begin abstract}{\begin{abstract}}}%
%BeginExpansion
\begin{abstract}%
%EndExpansion

A Han-Kobayashi based achievable scheme is presented for ergodic fading
two-user Gaussian interference channels (IFCs) with perfect channel state
information at all nodes and Gaussian codebooks with no time-sharing. Using
max-min optimization techniques, it is shown that jointly coding across all
states performs at least as well as separable coding for the sub-classes of
uniformly weak (every sub-channel is weak) and hybrid (mix of strong and weak
sub-channels that do not achieve the interference-free sum-capacity) IFCs. For
the uniformly weak IFCs, sufficient conditions are obtained for which the
sum-rate is maximized when interference is ignored at both receivers.%

%TCIMACRO{\TeXButton{End abstract}{\end{abstract}}}%
%BeginExpansion
\end{abstract}%
%EndExpansion

\section{Introduction}

Gaussian interference channels (IFCs) model wireless networks as a collection
of two or more interfering transmit-receive pairs (links). Capacity results
for two-user non-fading Gaussian\ IFCs are only known for specific sub-classes
of IFCs such as strong
\cite{cap_theorems:Sato_IC,cap_theorems:KobayashiHan_IC}, very strong (a
sub-class of strong) \cite{cap_theorems:Carleial_VSIFC}, one-sided weak
\cite{cap_theorems:Costa_IC}, and very weak or noisy
\cite{cap_theorems:ShangKramerChen}, \cite{cap_theorems:MotaKhan}, and
\cite{cap_theorems:AR_VVV}. Outer bounds for IFCs are developed in
\cite{cap_theorems:Kramer_IFCOB} and \cite{cap_theorems:ETW}. The best known
inner bounds are due to Han and Kobayashi (HK)
\cite{cap_theorems:KobayashiHan_IC}.

Ergodic fading and parallel Gaussian\ IFCs (PGICs) are IFC models that include
both the fading and interference characteristics of the wireless medium. PGICs
in which every sub-channel is strong and one-sided PGICs are studied in
\cite{cap_theorems:ChuCioffi_IC} and \cite{cap_theorems:SumCap_ParZIFC},
respectively; both papers present achievable schemes based on coding
independently for each parallel sub-channel. For PGICs,
\cite{cap_theorems:Shang_03} determines the conditions on the channel
coefficients and power constraints for which independent transmission across
sub-channels and treating interference as noise is optimal.

\begin{figure}
[ptb]
\begin{center}
\includegraphics[
height=2.5486in,
width=2.3134in
]%
{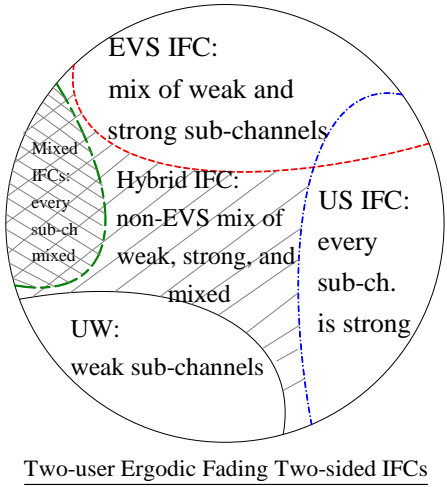}%
\caption{A Venn diagram representation of the four sub-classes of ergodic
fading IFCs.}%
\label{FigIFCVenn}%
\end{center}
\end{figure}

For ergodic fading Gaussian\ IFCs, henceforth referred to simply as IFCs, we
developed the sum-capacity and separability for specific sub-classes in
\cite{cap_theorems:SEP} and \cite{cap_theorems:SXEP}. In contrast to the
non-fading case, we proved that ergodic fading IFCs with a mix of weak and
strong sub-channels that satisfy a specific set of conditions can achieve the
sum of the interference-free capacities of the two intended links; we
identified such channels as the sub-class of \textit{ergodic very strong}
(EVS) IFCs. For this sub-class, we showed that jointly coding across all
sub-channels (i.e., transmitting the same message in every sub-channel) and
requiring the receivers to decode the transmissions from both users achieves
the capacity region. Furthermore, in \cite{cap_theorems:SEP}, we outlined the
optimality of this achievable coding scheme for a sub-class of
\textit{uniformly strong} (US) IFCs in which every sub-channel is strong. The
US and EVS sub-classes overlap but in general are not the same (see Fig.
\ref{FigIFCVenn}). For one-sided \textit{uniformly weak} (UW) IFCs, in which
every sub-channel is weak, we conjectured the optimality of ignoring
interference and separable coding in \cite{cap_theorems:SXEP}. Converse proofs
for each of the above-mentioned sub-channels as well as for a sub-class of
\textit{uniformly} \textit{mixed} two-sided IFCs, comprised of two
complementary UW and US one-sided IFCs, is developed in
\cite{cap_theorems:LSXSEEVP}.

For one-sided and two-sided IFCs, \cite{cap_theorems:LSXSEEVP} identifies a
sub-class of \textit{hybrid} IFCs (which is complementary to all previously
identified sub-classes; see Fig. \ref{FigIFCVenn}) comprised of a mix of
strong and weak sub-channels or a mix of strong, weak, and mixed sub-channels,
respectively, for which the EVS conditions are not satisfied. Specifically,
for one-sided IFCs, \cite{cap_theorems:LSXSEEVP} unifies the above-mentioned
capacity results using a HK-based achievable scheme that uses joint coding and
no time-sharing such that across all sub-channels, the interfering transmitter
sends a common and a private message. For the hybrid sub-class, this HK-based
scheme is shown to achieve a sum-rate at least as large as that achieved by
separable coding in which only common and private messages are sent in the
strong and weak sub-channels, respectively.

In this paper, we develop the HK-based achievable scheme using joint coding
for two-sided IFCs. We demonstrate the optimality of transmitting only common
messages for the two-sided EVS and US IFCs. The sum-capacity of two-sided UW
IFCs remains open; for the proposed HK-based joint coding scheme we determine
a set of sufficient conditions for which ignoring interference at both
receivers and separable coding maximize the sum-rate. Finally, we show that in
general for both the two-sided weak and the hybrid sub-classes, joint coding
of both private and common messages across all sub-channels achieves at least
as large a sum-rate as separable coding. Two-sided ergodic fading Gaussian
IFCs are studied in \cite{cap_theorems:Tuninetti} using a simplified form of
the HK\ region to determine the power policies that maximize a sum-rate inner
bound. In contrast, we focus on the problem of separability and use a max-min
optimization technique to unify known and new results for all sub-classes. The
paper is organized as follows. In\ Section \ref{Sec_II}, we present the
channel models studied. In Section \ref{Sec_III}, we summarize our main
results. We conclude in Section \ref{Sec_IV}.%

%TCIMACRO{\FRAME{ftbpFU}{2.3134in}{2.5486in}{0pt}{\Qcb{A Venn diagram
%representation of the four sub-classes of ergodic fading IFCs.}}%
%{\Qlb{FigIFCVenn}}{two_sided_ifcs_venn_noresults.eps}%
%{\special{ language "Scientific Word";  type "GRAPHIC";  display "USEDEF";
%valid_file "F";  width 2.3134in;  height 2.5486in;  depth 0pt;
%original-width 5.8046in;  original-height 6.0027in;  cropleft "0";
%filename 'Two_sided_IFCs_Venn_noresults.eps';file-properties "XNPEU";}}}%
%BeginExpansion
%EndExpansion

\section{\label{Sec_II}Channel Model}

A two-user ergodic fading Gaussian IFC consists of two transmit-receive pairs,
each pair indexed by $k$, for $k=1,2.$ In each use of the channel, transmitter
$k$ transmits the signal $X_{k}$ while receiver $k$ receives $Y_{k}$,
$k\in\mathcal{K}.$ For $\mathbf{X}=\left[  X_{1}\text{ }X_{2}\right]  ^{T}$,
the channel output vector $\mathbf{Y}=\left[  Y_{1}\text{ }Y_{2}\right]  ^{T}$
is given by%
\begin{equation}
\mathbf{Y}=\mathbf{HX}+\mathbf{Z} \label{IC_Y}%
\end{equation}
where $\mathbf{Z}=\left[  Z_{1}\text{ }Z_{2}\right]  ^{T}$ is a noise vector
with entries that are zero-mean, unit variance, circularly symmetric complex
Gaussian noise variables and $\mathbf{H}$ is a random matrix of fading gains
with entries $H_{m,k}$, for all $m,k\in\left\{  1,2\right\}  $, such that
$H_{m,k}$ denotes the fading gain between receiver $m$ and transmitter $k$. We
assume the fading process $\left\{  \mathbf{H}\right\}  $ is stationary and
ergodic but not necessarily Gaussian. Note that the channel gains $H_{m,k}$,
for all $m$ and $k$, are not assumed to be independent; however, $\mathbf{H}$
is known instantaneously at all the transmitters and receivers. A one-sided
fading Gaussian IFC results when either $H_{1,2}=0$ or $H_{2,1}=0$. A
two-sided IFC can be viewed as a collection of two complementary one-sided
IFCs, one with $H_{1,2}=0$ and the other with $H_{2,1}=0$.

Over $n$ uses of the channel, the transmit sequences $\left\{  X_{k,i}%
\right\}  $ are constrained in power according to%
\begin{equation}
\left.  \sum\limits_{i=1}^{n}\left\vert X_{k,i}\right\vert ^{2}\leq
n\overline{P}_{k}\right.  ,\text{ for all }k=1,2\text{.} \label{IFC_Pwr}%
\end{equation}
Since the transmitters know the fading states of the links on which they
transmit, they can allocate their transmitted signal powers according to the
channel state information. We write \underline{$P$}$(\mathbf{H})$ with entries
$P_{k}(\mathbf{H})$ for all $k$ to explicitly describe the power policy for
the entire set of random fading states. For an ergodic fading channel,
(\ref{IFC_Pwr}) then simplifies to
\begin{equation}
\left.  \mathbb{E}\left[  P_{k}(\mathbf{H})\right]  \leq\overline{P}%
_{k}\right.  \text{ for all }k=1,2, \label{ErgPwr}%
\end{equation}
where the expectation in (\ref{ErgPwr}) is over the distribution of
$\mathbf{H}$. We denote the set of all feasible policies $\underline{P}\left(
\mathbf{h}\right)  $, i.e., the power policies whose entries satisfy
(\ref{ErgPwr}), by $\mathcal{P}$.

Our definitions of average error probabilities, capacity regions, and
achievable rate pairs $\left(  R_{1},R_{2}\right)  $ for the IFC mirror the
standard information-theoretic definitions. Throughout the sequel, we use the
terms fading states and sub-channels interchangeably and refer to the ergodic
fading IFC as simply IFC. $\mathbb{E}\left[  \cdot\right]  $ denotes
expectation and $C(x)$ denotes $\log(1+x)$ where the logarithm is to the base 2.\ 

\section{\label{Sec_III}Achievable Scheme}

We consider an HK-based achievable scheme using Gaussian codebooks without
time-sharing and joint encoding and decoding across all sub-channels. We seek
to determine the power fractions allocated to private and common messages at
each transmitter that maximizes the sum-rate. Our motivation stems from the
fact that joint coding is optimal for EVS and US IFCs \cite{cap_theorems:SEP}
and achieves at least as large a sum-rate as separable coding for hybrid
one-sided IFCs \cite{cap_theorems:LSXSEEVP}. We outline the achievable scheme below.

Thus, transmitter $k$ transmits the same message pair $\left(  w_{kc}%
,w_{kp}\right)  $ in every sub-channel where $w_{kc}$ and $w_{kp}$ are the
common and private messages respectively. Each receiver decodes by jointly
decoding using the received signals from all sub-channels. Let $\alpha
_{k,\mathbf{H}}\in\left[  0,1\right]  $ and $\overline{\alpha}_{k,\mathbf{H}%
}=1-\alpha_{k,\mathbf{H}}$ denote the power fractions at transmitter $k$
allocated to transmitting the private and common messages, respectively, in
sub-channel $\mathbf{H}$. The two transmitted signals in each use of
sub-channel $\mathbf{H}$ are
\begin{subequations}
\label{HK2_signals}%
\begin{align}
X_{1}\left(  \mathbf{H}\right)   &  =\sqrt{\alpha_{1,\mathbf{H}}P_{1}\left(
\mathbf{H}\right)  }V_{1\mathbf{H}}+\sqrt{\overline{\alpha}_{1,\mathbf{H}%
}P_{1}\left(  \mathbf{H}\right)  }U_{1\mathbf{H}}\\
X_{2}\left(  \mathbf{H}\right)   &  =\sqrt{\alpha_{2,\mathbf{H}}P_{2}\left(
\mathbf{H}\right)  }V_{2\mathbf{H}}+\sqrt{\overline{\alpha}_{2,\mathbf{H}%
}P_{2}\left(  \mathbf{H}\right)  }U_{2\mathbf{H}}%
\end{align}
where $V_{k\mathbf{H}}$ and $U_{k\mathbf{H}}$, $k=1,2,$ are independent
zero-mean unit variance Gaussian random variables, for all $\mathbf{H}$. We
use the notation $V_{k\mathbf{H}}$ and $U_{k\mathbf{H}}$ to indicate that the
random variables are mutually independent for every instantiation of
$\mathbf{H}$, i.e., \ independent codebooks in each sub-channel. Let
$\underline{\alpha}_{\mathbf{H}}$ denote a vector of power fractions with
entries $\alpha_{k,\mathbf{H}}$, $k=1,2.$

In \cite[Theorem 4]{cap_theorems:CMG_ITA}, the authors present the HK region
achieved by superposition coding. Assuming no time-sharing, for the Gaussian
signaling in (\ref{HK2_signals}) and with joint coding, one can directly
extend the analysis in \cite[Theorem 4]{cap_theorems:CMG_ITA} to ergodic
fading Gaussian IFCs. The following Proposition based on \cite[Theorem
4]{cap_theorems:CMG_ITA} summarizes the resulting rate bounds.
\end{subequations}
\begin{proposition}
A rate pair $\left(  R_{1},R_{2}\right)  $ is achievable for a HK scheme with
superposition coding and no time-sharing for ergodic fading IFCs if
\begin{align}
R_{k}  &  \leq B_{k}\left(  \underline{\alpha}_{\mathbf{H}},\underline
{P}\left(  \mathbf{H}\right)  \right) \\
R_{1}+R_{2}  &  \leq B_{3}\left(  \underline{\alpha}_{\mathbf{H}}%
,\underline{P}\left(  \mathbf{H}\right)  \right) \\
R_{1}+R_{2}  &  \leq B_{4}\left(  \underline{\alpha}_{\mathbf{H}}%
,\underline{P}\left(  \mathbf{H}\right)  \right) \\
R_{1}+R_{2}  &  \leq B_{5}\left(  \underline{\alpha}_{\mathbf{H}}%
,\underline{P}\left(  \mathbf{H}\right)  \right) \\
2R_{1}+R_{2}  &  \leq B_{6}\left(  \underline{\alpha}_{\mathbf{H}}%
,\underline{P}\left(  \mathbf{H}\right)  \right) \\
R_{1}+2R_{2}  &  \leq B_{7}\left(  \underline{\alpha}_{\mathbf{H}}%
,\underline{P}\left(  \mathbf{H}\right)  \right)
\end{align}
where%
\begin{equation}
B_{k}=%
\begin{array}
[c]{cc}%
\mathbb{E}\left[  C\left(  \frac{\left\vert H_{k,k}\right\vert ^{2}%
P_{k}\left(  \mathbf{H}\right)  }{1+\alpha_{j,\mathbf{H}}\left\vert
H_{k,j}\right\vert ^{2}P_{j}\left(  \mathbf{H}\right)  }\right)  \right]  , &
j,k=1,2,j\not =k
\end{array}
\end{equation}%
\begin{align}
B_{3}  &  =\mathbb{E}\left[  C\left(  \frac{\left\vert H_{1,1}\right\vert
^{2}P_{1}\left(  \mathbf{H}\right)  +\left\vert H_{1,2}\right\vert
^{2}\overline{\alpha}_{2,\mathbf{H}}P_{2}\left(  \mathbf{H}\right)  }%
{1+\alpha_{2,\mathbf{H}}\left\vert H_{1,2}\right\vert ^{2}P_{2}\left(
\mathbf{H}\right)  }\right)  \right] \label{B3_def}\\
&  +\mathbb{E}\left[  C\left(  \frac{\left\vert H_{2,2}\right\vert ^{2}%
\alpha_{2,\mathbf{H}}P_{2}\left(  \mathbf{H}\right)  }{1+\alpha_{1,\mathbf{H}%
}\left\vert H_{2,1}\right\vert ^{2}P_{1}\left(  \mathbf{H}\right)  }\right)
\right] \nonumber\\
B_{4}  &  =\left.  B_{3}\right\vert _{\text{indices 1 and 2 swapped}}%
\end{align}%
\begin{align}
B_{5}  &  =\mathbb{E}\left[  C\left(  \frac{\alpha_{1,\mathbf{H}}\left\vert
H_{1,1}\right\vert ^{2}P_{1}\left(  \mathbf{H}\right)  +\left\vert
H_{1,2}\right\vert ^{2}\overline{\alpha}_{2,\mathbf{H}}P_{2}\left(
\mathbf{H}\right)  }{1+\alpha_{2,\mathbf{H}}\left\vert H_{1,2}\right\vert
^{2}P_{2}\left(  \mathbf{H}\right)  }\right)  \right] \\
&  +\mathbb{E}\left[  C\left(  \frac{\alpha_{2,\mathbf{H}}\left\vert
H_{2,2}\right\vert ^{2}P_{2}\left(  \mathbf{H}\right)  +\left\vert
H_{2,1}\right\vert ^{2}\overline{\alpha}_{1,\mathbf{H}}P_{1}\left(
\mathbf{H}\right)  }{1+\alpha_{1,\mathbf{H}}\left\vert H_{2,1}\right\vert
^{2}P_{1}\left(  \mathbf{H}\right)  }\right)  \right] \nonumber
\end{align}%
\begin{align}
B_{6}  &  =\mathbb{E}\left[  C\left(  \frac{\left\vert H_{1,1}\right\vert
^{2}P_{1}\left(  \mathbf{H}\right)  +\left\vert H_{1,2}\right\vert
^{2}\overline{\alpha}_{2,\mathbf{H}}P_{2}\left(  \mathbf{H}\right)  }%
{1+\alpha_{2,\mathbf{H}}\left\vert H_{1,2}\right\vert ^{2}P_{2}\left(
\mathbf{H}\right)  }\right)  \right] \\
&  +\mathbb{E}\left[  C\left(  \frac{\left\vert H_{1,1}\right\vert ^{2}%
\alpha_{1,\mathbf{H}}P_{1}\left(  \mathbf{H}\right)  }{1+\alpha_{2,\mathbf{H}%
}\left\vert H_{1,2}\right\vert ^{2}P_{2}\left(  \mathbf{H}\right)  }\right)
\right] \nonumber\\
&  +\mathbb{E}\left[  C\left(  \frac{\alpha_{2,\mathbf{H}}\left\vert
H_{2,2}\right\vert ^{2}P_{1}\left(  \mathbf{H}\right)  +\left\vert
H_{2,1}\right\vert ^{2}\overline{\alpha}_{1,\mathbf{H}}P_{1}\left(
\mathbf{H}\right)  }{1+\alpha_{1,\mathbf{H}}\left\vert H_{2,1}\right\vert
^{2}P_{1}\left(  \mathbf{H}\right)  }\right)  \right] \nonumber
\end{align}%
\begin{equation}
B_{7}=\left.  B_{6}\right\vert _{\text{indices 1 and 2 swapped}}%
\end{equation}

\end{proposition}

\begin{theorem}
The sum-capacity of ergodic fading IFCs is lower bounded by%
\begin{equation}
\max_{\underline{P}\left(  \mathbf{H}\right)  \in\mathcal{P},\alpha
_{k,\mathbf{H}}\in\lbrack0,1]}\min_{m\in\left\{  1,2,3,4,5,6\right\}  }%
S_{m}\left(  \underline{\alpha}_{\mathbf{H}},\underline{P}\left(
\mathbf{H}\right)  \right)  \label{HK2_SR}%
\end{equation}
where
\begin{align}
S_{1}\left(  \underline{\alpha}_{\mathbf{H}},\underline{P}\left(
\mathbf{H}\right)  \right)   &  =B_{1}\left(  \underline{\alpha}_{\mathbf{H}%
},\underline{P}\left(  \mathbf{H}\right)  \right)  +B_{2}\left(
\underline{\alpha}_{\mathbf{H}},\underline{P}\left(  \mathbf{H}\right)
\right)  ,\\
S_{j}\left(  \underline{\alpha}_{\mathbf{H}},\underline{P}\left(
\mathbf{H}\right)  \right)   &  =B_{j+1}\left(  \underline{\alpha}%
_{\mathbf{H}},\underline{P}\left(  \mathbf{H}\right)  \right)  ,\text{
}j=2,3,4\label{S2_def}\\
S_{5}\left(  \underline{\alpha}_{\mathbf{H}},\underline{P}\left(
\mathbf{H}\right)  \right)   &  =\left(  B_{6}\left(  \underline{\alpha
}_{\mathbf{H}},\underline{P}\left(  \mathbf{H}\right)  \right)  +B_{2}\left(
\underline{\alpha}_{\mathbf{H}},\underline{P}\left(  \mathbf{H}\right)
\right)  \right)  /2\\
S_{6}\left(  \underline{\alpha}_{\mathbf{H}},\underline{P}\left(
\mathbf{H}\right)  \right)   &  =\left(  B_{7}\left(  \underline{\alpha
}_{\mathbf{H}},\underline{P}\left(  \mathbf{H}\right)  \right)  +B_{1}\left(
\underline{\alpha}_{\mathbf{H}},\underline{P}\left(  \mathbf{H}\right)
\right)  \right)  /2.
\end{align}
\textit{(a)} For EVS\ IFCs, the sum-capacity $S_{1}\left(  \underline
{0},\underline{P}^{(wf)}\left(  \mathbf{H}\right)  \right)  $ is achieved by
choosing \underline{$\alpha$}$_{\mathbf{H}}^{\ast}=\underline{0}$ for all
$\mathbf{H}$ and $\underline{P}^{\ast}\left(  \mathbf{H}\right)
=\underline{P}^{\left(  wf\right)  }\left(  \mathbf{H}\right)  $ provided
$S_{1}\left(  \underline{0},\underline{P}^{(wf)}\left(  \mathbf{H}\right)
\right)  <S_{j}\left(  \underline{0},\underline{P}^{(wf)}\left(
\mathbf{H}\right)  \right)  $, for all $j>1$, where $\underline{P}^{\left(
wf\right)  }\left(  \mathbf{H}\right)  $ is the optimal waterfilling policy
for the two interference-free direct links. \textit{(b)} For US IFCs, the
sum-capacity is achieved by $\underline{\alpha}_{\mathbf{H}}^{\ast}=0$, for
all $\mathbf{H}$ and is given by
\begin{equation}
\max_{\underline{P}\left(  \mathbf{H}\right)  \in\mathcal{P}}\min
_{m\in\left\{  1,2,3\right\}  }S_{m}\left(  \underline{0},\underline{P}\left(
\mathbf{H}\right)  \right)  . \label{US_SC}%
\end{equation}
\textit{(c)} For UM\ IFCs, the sum-capacity is achieved by choosing
$\alpha_{k,\mathbf{H}}^{\ast}=1$ and $\alpha_{j,\mathbf{H}}^{\ast}=0$,
$j\not =k$, where $k$ and $j$ are the receivers that see weak and strong
interference, respectively, and is given by
\begin{equation}
\max_{\underline{P}\left(  \mathbf{H}\right)  \in\mathcal{P}}\min
_{m\in\left\{  2,3\right\}  }S_{m}\left(  \underline{\alpha}_{\mathbf{H}%
}^{\ast},\underline{P}\left(  \mathbf{H}\right)  \right)  . \label{UM_SC}%
\end{equation}
\textit{(d)} For UW IFCs, the sum-rate is maximized by $\underline{\alpha
}_{\mathbf{H}}^{\ast}=\underline{1}$ if, for every $\underline{P}\left(
\mathbf{H}\right)  \in\mathcal{P}$,%
\begin{align}
\left\vert H_{2,2}\right\vert ^{2}  &  >\left(  1+\left\vert H_{2,1}%
\right\vert ^{2}P_{1}\left(  \mathbf{H}\right)  \right)  \left\vert
H_{1,2}\right\vert ^{2}\label{UW_C1}\\
\left\vert H_{1,1}\right\vert ^{2}  &  >\left(  1+\left\vert H_{1,2}%
\right\vert ^{2}P_{2}\left(  \mathbf{H}\right)  \right)  \left\vert
H_{2,1}\right\vert ^{2} \label{UW_C2}%
\end{align}
and is given by
\begin{equation}
\max_{\underline{P}\left(  \mathbf{H}\right)  \in\mathcal{P}}S_{1}\left(
\underline{1},\underline{P}\left(  \mathbf{H}\right)  \right)  . \label{UW_SR}%
\end{equation}
For a hybrid one-sided IFC, the achievable sum-rate is maximized by
\begin{equation}
\alpha_{k,\mathbf{H}}^{\ast}=\left\{
\begin{array}
[c]{cc}%
\alpha_{k}\left(  \mathbf{H}\right)  \in(0,1] & \mathbf{H}\text{ is weak }\\
0 & \mathbf{H}\text{ is strong.}%
\end{array}
\right.  ,k=1,2,
\end{equation}
and is given by (\ref{HK2_SR}) for this choice of $\underline{\alpha
}_{\mathbf{H}}^{\ast}$.
\end{theorem}

\begin{remark}
The conditions in (\ref{UW_C1}) and (\ref{UW_C2}) hold for all feasible power
policies $\underline{P}\left(  \mathbf{H}\right)  $, i.e., policies satisfying
the fading averaged constraint in (\ref{ErgPwr}), and thus, are quite
restrictive in defining the set of channel gains for which ignoring
interference is optimal for UW IFCs. However, the analysis and the conditions
(\ref{UW_C1}) and (\ref{UW_C2}) also hold for ergodic channels with a
per-symbol or equivalently per-fading state power constraint for which
determining the largest values of the right-side of (\ref{UW_C1}) and
(\ref{UW_C2}) is relatively easier.
\end{remark}

\begin{proof}
Our proof relies on using the fact that the maximization of the minimum of two
functions, say $f_{1}\left(  \alpha_{\mathbf{H}},\underline{P}\left(
\mathbf{H}\right)  \right)  $ and $f_{2}\left(  \alpha_{\mathbf{H}}%
,\underline{P}\left(  \mathbf{H}\right)  \right)  $ is equivalent to a
\textit{minimax }optimization problem (see for e.g., \cite[II.C]%
{cap_theorems:HVPoor01}) for which the maximum sum-rate $S^{\ast}$ is given by
the following three cases. In each case, the optimal $\underline{P}^{\ast
}\left(  \mathbf{H}\right)  $ and $\alpha_{\mathbf{H}}^{\ast}$ maximize the
smaller of the two functions and therefore maximize both for the case when the
two functions are equal. The three cases are
\begin{subequations}
\label{Hyb_MM}%
\begin{align}
&
\begin{array}
[c]{cc}%
\text{Case }1: & S^{\ast}=f_{1}\left(  \alpha_{\mathbf{H}}^{\ast}%
,\underline{P}^{\ast}\left(  \mathbf{H}\right)  \right)  <f_{2}\left(
\alpha_{\mathbf{H}}^{\ast},\underline{P}^{\ast}\left(  \mathbf{H}\right)
\right)
\end{array}
\label{MMC1}\\
&
\begin{array}
[c]{cc}%
\text{Case }2: & S^{\ast}=f_{2}\left(  \alpha_{\mathbf{H}}^{\ast}%
,\underline{P}^{\ast}\left(  \mathbf{H}\right)  \right)  <f_{1}\left(
\alpha_{\mathbf{H}}^{\ast},\underline{P}^{\ast}\left(  \mathbf{H}\right)
\right)
\end{array}
\label{MMC2}\\
&
\begin{array}
[c]{cc}%
\text{Case }3: & S^{\ast}=f_{1}\left(  \alpha_{\mathbf{H}}^{\ast}%
,\underline{P}^{\ast}\left(  \mathbf{H}\right)  \right)  =f_{2}\left(
\alpha_{\mathbf{H}}^{\ast},\underline{P}^{\ast}\left(  \mathbf{H}\right)
\right)
\end{array}
\label{MMC3}%
\end{align}
From (\ref{HK2_SR}), the sum-rate is the solution to a max-min optimization of
$f_{1}\left(  \cdot\right)  =S_{1}\left(  \cdot\right)  $ and $f_{2}\left(
\cdot\right)  =\min_{j>1}S_{j}\left(  \cdot\right)  $. We now consider each
sub-class separately.

\textit{Ergodic very\ strong}: By definition, an EVS\ IFC results when the sum
of the interference-free capacities of the two links can be achieved. From
(\ref{Hyb_MM}), one special case of the max-min optimization in (\ref{HK2_SR})
corresponds to the EVS\ sub-class. This results when
\end{subequations}
\begin{multline}
\max_{\underline{P}\left(  \mathbf{H}\right)  ,\alpha_{\mathbf{H}}}%
S_{1}\left(  \underline{\alpha}_{\mathbf{H}},\underline{P}\left(
\mathbf{H}\right)  \right)  =S_{1}\left(  \underline{0},\underline{P}^{\left(
wf\right)  }\left(  \mathbf{H}\right)  \right) \label{EVS2}\\
<S_{j}\left(  \underline{0},\underline{P}^{\left(  wf\right)  }\left(
\mathbf{H}\right)  \right)  ,\text{ for all }j>1\text{,}%
\end{multline}
where we have used the fact that the ergodic capacities of the two
interference-free links are maximized by the optimal single-user waterfilling
policies \cite{cap_theorems:GoldsmithVaraiya}, denoted by $\underline
{P}^{\left(  wf\right)  }\left(  \mathbf{H}\right)  $ with entries
$P_{k}^{\left(  wf\right)  }\left(  H_{k,k}\right)  $. Note that $S_{2}\left(
\underline{0},\underline{P}\left(  \mathbf{H}\right)  \right)  $ and
$S_{3}\left(  \underline{0},\underline{P}\left(  \mathbf{H}\right)  \right)  $
are the multiple-access sum-capacities at receivers 1 and 2, respectively,
such that%
\begin{equation}
S_{2}\left(  \underline{0},\underline{P}\left(  \mathbf{H}\right)  \right)
=\left.  S_{3}\left(  \underline{0},\underline{P}\left(  \mathbf{H}\right)
\right)  \right\vert _{\text{swap indices 1 and 2}}.
\end{equation}
We now show that (\ref{EVS2}) simplifies to the requirement
\begin{multline}
S_{1}\left(  \underline{0},\underline{P}^{\left(  wf\right)  }\left(
\mathbf{H}\right)  \right)  <\min\left(  S_{2}\left(  \underline{0}%
,\underline{P}^{\left(  wf\right)  }\left(  \mathbf{H}\right)  \right)
,\right. \label{EVS_C}\\
\left.  S_{3}\left(  \underline{0},\underline{P}^{\left(  wf\right)  }\left(
\mathbf{H}\right)  \right)  \right)
\end{multline}
The proof follows trivially by expanding the terms $S_{j}\left(  \underline
{0},\underline{P}^{\left(  wf\right)  }\left(  \mathbf{H}\right)  \right)  $
for all $j>1$ and comparing them to $S_{1}\left(  \underline{0},\underline
{P}^{\left(  wf\right)  }\left(  \mathbf{H}\right)  \right)  $. We illustrate
this for $S_{4}\left(  \underline{0},\underline{P}\left(  \mathbf{H}\right)
\right)  $ as follows.
\begin{subequations}
\label{EVS_S}%
\begin{align}
S_{4}\left(  \underline{0},\underline{P}\left(  \mathbf{H}\right)  \right)
&  =\mathbb{E}\left[  C\left(  \left\vert H_{1,2}\right\vert ^{2}%
P_{2}^{\left(  wf\right)  }\left(  \mathbf{H}\right)  \right)  \right] \\
&  \text{ \ \ \ }+\mathbb{E}\left[  C\left(  \left\vert H_{2,1}\right\vert
^{2}P_{1}^{\left(  wf\right)  }\left(  \mathbf{H}\right)  \right)  \right]
\nonumber\\
&  >\text{ }\mathbb{E}\left[  C\left(  \left\vert H_{2,2}\right\vert ^{2}%
P_{2}^{\left(  wf\right)  }\left(  \mathbf{H}\right)  \right)  \right]
\label{EVS_S2}\\
&  \text{ \ \ \ }+\mathbb{E}\left[  C\left(  \left\vert H_{2,1}\right\vert
^{2}P_{1}^{\left(  wf\right)  }\left(  \mathbf{H}\right)  \right)  \right]
\nonumber\\
&  \geq S_{3}\left(  \underline{0},\underline{P}^{\left(  wf\right)  }\left(
\mathbf{H}\right)  \right)  \label{EVS_S3}%
\end{align}
where (\ref{EVS_S2}) follows from simplifying (\ref{EVS_C}) by expanding the
multiple-access sum-capacity terms and (\ref{EVS_S2}) follows from using chain
rule for mutual information. We note that (\ref{EVS_C}) is the EVS condition
developed in \cite[Th. 7]{cap_theorems:SEP} (see also \cite[Theorem
2]{cap_theorems:LSXSEEVP}). The sum-capacity follows from noting that the sum
of the capacities of two interference-free links is an outer bound on the
IFC\ sum-capacity.%
%TCIMACRO{\FRAME{ftbpFU}{2.6238in}{2.6238in}{0pt}{\Qcb{Two-sided ergodic fading
%IFCs: overview of known results.}}{\Qlb{Fig_IFCVenn}}{two_sided_ifcs_venn.eps}%
%{\special{ language "Scientific Word";  type "GRAPHIC";  display "USEDEF";
%valid_file "F";  width 2.6238in;  height 2.6238in;  depth 0pt;
%original-width 6.0424in;  original-height 5.6239in;  cropleft "0";
%filename 'Two_sided_IFCs_Venn.eps';file-properties "XNPEU";}}}%
%BeginExpansion
\begin{figure}
[ptb]
\begin{center}
\includegraphics[
height=2.6238in,
width=2.6238in
]%
{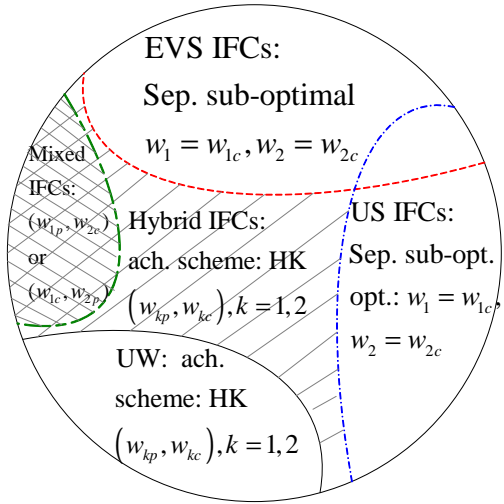}%
\caption{Two-sided ergodic fading IFCs: overview of known results.}%
\label{Fig_IFCVenn}%
\end{center}
\end{figure}
%EndExpansion

\textit{Uniformly strong}: One can verify in a straightforward manner that
$S_{1}\left(  \alpha_{\mathbf{H}},\underline{P}\left(  \mathbf{H}\right)
\right)  $ is maximized $\alpha_{1,\mathbf{H}}^{\ast}=\alpha_{2,\mathbf{H}%
}^{\ast}=0$. Furthermore, when all sub-channels are strong, i.e., when
$\Pr[\left\vert H_{1,2}\right\vert >\left\vert H_{2,2}\right\vert ]=1,$ the
bound $S_{2}\left(  \alpha_{\mathbf{H}},\underline{P}\left(  \mathbf{H}%
\right)  \right)  $ in (\ref{S2_def}) can be rewritten as
\end{subequations}
\begin{multline}
\mathbb{E}\left[  C\left(  \left\vert H_{1,1}\right\vert ^{2}P_{1}\left(
\mathbf{H}\right)  +\left\vert H_{1,2}\right\vert ^{2}P_{2}\left(
\mathbf{H}\right)  \right)  \right] \label{US1}\\
-\mathbb{E}\left[  C\left(  1+\alpha_{2,\mathbf{H}}\left\vert H_{1,2}%
\right\vert ^{2}P_{2}\left(  \mathbf{H}\right)  \right)  \right] \\
+\mathbb{E}\left[  C\left(  \frac{\left\vert H_{2,2}\right\vert ^{2}%
\alpha_{2,\mathbf{H}}P_{2}\left(  \mathbf{H}\right)  }{1+\alpha_{1,\mathbf{H}%
}\left\vert H_{2,1}\right\vert ^{2}P_{1}\left(  \mathbf{H}\right)  }\right)
\right]  .
\end{multline}
Using the US condition, one can verify that for every choice of $\underline
{P}\left(  \mathbf{H}\right)  $, $S_{2}\left(  \alpha_{\mathbf{H}}%
,\underline{P}\left(  \mathbf{H}\right)  \right)  $ is maximized by
$\alpha_{1,\mathbf{H}}^{\ast}=\alpha_{2,\mathbf{H}}^{\ast}=0$, i.e.,
$w_{k}=w_{k,c}$, $k=1,2$. Since $S_{3}\left(  \cdot\right)  $ is obtained from
$S_{2}\left(  \cdot\right)  $ by swapping the indices, the above choice also
maximizes $S_{3}\left(  \cdot\right)  $. Transmitting only common messages at
both transmitters results in multiple-access regions at both receivers; one
can use the properties of these multiple-access regions to show that the
remaining sum-rate bounds are at least as much as the minimum of
$S\,_{j}\left(  \underline{0},\cdot\right)  ,$ $j=1,2,3,$ such that the
maximum achievable sum-rate is given by (\ref{US_SC}). The outer bound
analysis in \cite[Theorem 3]{cap_theorems:LSXSEEVP} helps establish that
(\ref{US_SC}) is the US sum-capacity.

\textit{Uniformly mixed}: Without loss of generality, assume $\Pr[\left\vert
H_{2,1}\right\vert ^{2}>\left\vert H_{1,1}\right\vert ^{2}]=1$ and
$\Pr[\left\vert H_{1,2}\right\vert ^{2}<\left\vert H_{2,2}\right\vert
^{2}]=1,$ i.e., receivers 1 and 2 experience weak and strong interference,
respectively. Comparing with the US case, we choose $\alpha_{1,\mathbf{H}%
}^{\ast}=0$. Furthermore, is is straightforward to verify that $S_{2}\left(
\underline{\alpha}_{\mathbf{H}},\underline{P}\left(  \mathbf{H}\right)
\right)  $ is maximized by $\alpha_{2,\mathbf{H}}^{\ast}=1$ while $S_{3}$ is
independent of $\alpha_{1,\mathbf{H}}$ for $\alpha_{1,\mathbf{H}}^{\ast}=0$.
For $j\geq4$, $S_{j}\left(  \alpha_{1,\mathbf{H}}^{\ast}=0,\alpha
_{2,\mathbf{H}},\underline{P}\left(  \mathbf{H}\right)  \right)  $ is in
general maximized by a $\alpha_{2,\mathbf{H}}\not =1$. Evaluating all
functions at $\left(  \alpha_{1,\mathbf{H}}^{\ast},\alpha_{2,\mathbf{H}}%
^{\ast}\right)  =\left(  0,1\right)  $ and for any $\underline{P}\left(
\mathbf{H}\right)  $, one can verify that $S_{1}\left(  \cdot\right)
=S_{2}\left(  \cdot\right)  $, $S_{4}\left(  \cdot\right)  =S_{3}\left(
\cdot\right)  $, and $S_{j}\left(  \cdot\right)  >\min\left(  S_{2}\left(
\cdot\right)  ,S_{3}\left(  \cdot\right)  \right)  $, $j=5,6$. Thus, the
max-min optimization simplifies to (\ref{UM_SC}). Finally, using outer bounds
developed for two complementary one-sided IFCs (see \cite[Theorem
5]{cap_theorems:LSXSEEVP}), we can show that (\ref{UM_SC}) is the sum-capacity.

\textit{Uniformly weak}: For this sub-class of channels, it is straightforward
to see that the conditions for Case 1 in (\ref{EVS2}) will not be satisfied
(as otherwise the sub-class would be EVS), and thus, $\alpha_{k,\mathbf{H}%
}^{\ast}\not =0$ for $k=1,2$. For the \ case with one-sided interference, in
\cite[Th. 4]{cap_theorems:LSXSEEVP}, we show that transmitting only private
messages at the interfering transmitter maximizes the sum-rate and is in fact
sum-capacity optimal. However, for the two-sided case, the choice of
$\alpha_{k,\mathbf{H}}^{\ast}=1$ for all $k$, i.e., $w_{k}=w_{k,p}$ for all
$k$, does not necessarily maximize the sum-rate. Consider the function
$S_{2}\left(  \underline{\alpha}_{\mathbf{H}},\underline{P}\left(
\mathbf{H}\right)  \right)  $ in (\ref{S2_def}). From (\ref{B3_def}), it can
be rewritten as%
\begin{multline}
S_{2}\left(  \underline{\alpha}_{\mathbf{H}},\underline{P}\left(
\mathbf{H}\right)  \right)  =\mathbb{E}\left[  C\left(  \left\vert
H_{1,1}\right\vert ^{2}P_{1}\left(  \mathbf{H}\right)  +\left\vert
H_{1,2}\right\vert ^{2}P_{2}\left(  \mathbf{H}\right)  \right)  \right.
\label{S2_weak}\\
+C\left(  \left\vert H_{2,1}\right\vert ^{2}\alpha_{1,\mathbf{H}}P_{1}\left(
\mathbf{H}\right)  +\left\vert H_{2,2}\right\vert ^{2}\alpha_{2,\mathbf{H}%
}P_{2}\left(  \mathbf{H}\right)  \right) \\
-C\left(  1+\alpha_{2,\mathbf{H}}\left\vert H_{1,2}\right\vert ^{2}%
P_{2}\left(  \mathbf{H}\right)  \right)  -\left.  C\left(  1+\alpha
_{1,\mathbf{H}}\left\vert H_{2,1}\right\vert ^{2}P_{1}\left(  \mathbf{H}%
\right)  \right)  \right]  .
\end{multline}
For $\alpha_{1,\mathbf{H}}=0$, one can use the concavity of the logarithm to
show that $S_{2}\left(  \underline{\alpha}_{\mathbf{H}},\underline{P}\left(
\mathbf{H}\right)  \right)  $ is maximized by $\alpha_{2,\mathbf{H}}^{\ast}%
=1$; however, for any $\alpha_{1,\mathbf{H}}>0$ and a $P_{1}\left(
\mathbf{H}\right)  \not =0$, $S_{2}\left(  \cdot\right)  $ may not be
maximized by $\alpha_{2,\mathbf{H}}^{\ast}=1$. Rewriting the second and third
term in (\ref{S2_weak}) as
\begin{align*}
&  \mathbb{E}\left[  C\left(  \left\vert H_{2,1}\right\vert ^{2}%
\alpha_{1,\mathbf{H}}P_{1}\left(  \mathbf{H}\right)  \right)  \right.
+C\left(  \frac{\left\vert H_{2,2}\right\vert ^{2}\alpha_{2,\mathbf{H}}%
P_{2}\left(  \mathbf{H}\right)  }{1+\left\vert H_{2,1}\right\vert ^{2}%
\alpha_{1,\mathbf{H}}P_{1}\left(  \mathbf{H}\right)  }\right) \\
&  -\left.  C\left(  1+\alpha_{2,\mathbf{H}}\left\vert H_{1,2}\right\vert
^{2}P_{2}\left(  \mathbf{H}\right)  \right)  \right]
\end{align*}
we see that $\underline{\alpha}_{\mathbf{H}}^{\ast}=\underline{1}$ maximizes
$S_{2}$ only if (\ref{UW_C1}) is satisfied for all $\underline{P}\left(
\mathbf{H}\right)  $. One can similarly show that the choice $\underline
{\alpha}_{\mathbf{H}}^{\ast}=\underline{1}$ maximizes $S_{3}$ if (\ref{UW_C2})
is satisfied for all $\underline{P}\left(  \mathbf{H}\right)  $. Both
(\ref{UW_C1}) and (\ref{UW_C2}) need to be satisfied for $\underline{\alpha
}_{\mathbf{H}}=\underline{1}$ to maximize $S_{4}$. In general, (\ref{UW_C1})
and (\ref{UW_C2}) do not guarantee that $\underline{\alpha}_{\mathbf{H}}%
^{\ast}=\underline{1}$ maximizes $S_{1}$, $S_{5}$, and $S_{6}$; however, since
$S_{j}\left(  \underline{\alpha}_{\mathbf{H}}^{\ast}=\underline{1}%
,\underline{P}_{\mathbf{H}}\right)  =S_{2}\left(  \underline{\alpha
}_{\mathbf{H}}^{\ast}=\underline{1},\underline{P}_{\mathbf{H}}\right)  $,
$j=1,5,6$, the maximum sum-rate for the UW\ sub-class is given by
(\ref{UW_SR}) when (\ref{UW_C1}) and (\ref{UW_C2}) hold.

\textit{Hybrid}: This sub-class includes all IFCs with a mix of weak, strong,
and mixed sub-channels that to do not satisfy the EVS condition. Let
$\underline{\alpha}_{\mathbf{H}}^{\left(  s\right)  }$, $\underline{\alpha
}_{\mathbf{H}}^{\left(  m\right)  }$, and $\underline{\alpha}_{\mathbf{H}%
}^{\left(  w\right)  }$ denote the vector of power fractions for the private
messages in the strong, mixed, and weak sub-channels, respectively. Using the
linearity of expectation, we can write the expressions for $S_{j}\left(
\cdot\right)  $ for all $j$ as sums of expectations of the appropriate bounds
over the collection of strong, mixed, and weak sub-channels. Let
$S_{j}^{\left(  s\right)  }\left(  \cdot\right)  $ $S_{j}^{\left(  m\right)
}\left(  \cdot\right)  $, and $S_{j}^{\left(  w\right)  }\left(  \cdot\right)
$ denote the expectation over the strong, mixed, and weak sub-channels,
respectively, such that $S_{j}\left(  \cdot\right)  =S_{j}^{\left(  s\right)
}\left(  \cdot\right)  +S_{j}^{\left(  m\right)  }+S_{j}^{\left(  w\right)
}\left(  \cdot\right)  $, for all $j.$ Let $\underline{\alpha}_{\mathbf{H}%
}^{(s)}$, $\underline{\alpha}_{\mathbf{H}}^{(m)}$, and $\underline{\alpha
}_{\mathbf{H}}^{(w)}$ denote the optimal $\underline{\alpha}_{\mathbf{H}%
}^{\ast}$ for the weak and strong states, respectively. For those sub-channels
which are strong and mixed, one can use the arguments above to show that
$\underline{\alpha}_{\mathbf{H}}^{(s)}=0$ and $\underline{\alpha}_{\mathbf{H}%
}^{(m)}=\left(  0,1\right)  $ maximize $S_{j}^{\left(  s\right)  }\left(
\cdot\right)  $ and $S_{j}^{\left(  m\right)  }$, respectively. For the weak
sub-channels, as for the UW\ sub-class, the entries of the optimal
$\underline{\alpha}_{\mathbf{H}}^{(w)}$ can take on any value in the range
$(0,1]$.
\end{proof}

\section{\label{Sec_IV}Concluding Remarks}

We have presented a Han-Kobayashi based achievable scheme for two-sided
ergodic fading Gaussian IFCs. Relying on converse results in
\cite{cap_theorems:LSXSEEVP}, we have shown the optimality of transmitting
only common messages for the sub-classes of EVS and US IFCs. For the hybrid
sub-classes, we have shown that the proposed joint coding scheme does at least
as well as separable coding by exploiting the strong sub-channels to reduce
interference in the weak sub-channels. In contrast to the one-sided UW
sub-class for which ignoring interference and separable coding are optimal, we
have argued here that in general joint coding is required for the two-sided
UW\ sub-class. While the sum-capacity optimal scheme is unknown for this
sub-class, we have developed a set of sufficient conditions under which
ignoring interference and separable coding are optimal. Our results are
summarized by the Venn diagrams in Fig. \ref{Fig_IFCVenn}.

\bibliographystyle{IEEEtran}
\bibliography{IC_refs}

\end{document}